\documentclass[12pt]{article}
\usepackage{amssymb}
\renewcommand{\theequation}{\arabic{section}.\arabic{equation}}

\begin{document}



\def\a{\alpha}
\def\b{\beta}
\def\d{\delta}
\def\e{\epsilon}
\def\g{\gamma}
\def\h{\mathfrak{h}}
\def\k{\kappa}
\def\l{\lambda}
\def\o{\omega}
\def\p{\wp}
\def\r{\rho}
\def\t{\tau}
\def\s{\sigma}
\def\z{\zeta}
\def\x{\xi}
\def\V={{{\bf\rm{V}}}}
 \def\A{{\cal{A}}}
 \def\B{{\cal{B}}}
 \def\C{{\cal{C}}}
 \def\D{{\cal{D}}}
\def\G{\Gamma}
\def\K{{\cal{K}}}
\def\O{\Omega}
\def\R{\bar{R}}
\def\T{{\cal{T}}}
\def\L{\Lambda}
\def\f{E_{\tau,\eta}(sl_2)}
\def\E{E_{\tau,\eta}(sl_n)}
\def\Zb{\mathbb{Z}}
\def\Cb{\mathbb{C}}

\def\R{\overline{R}}

\def\beq{\begin{equation}}
\def\eeq{\end{equation}}
\def\bea{\begin{eqnarray}}
\def\eea{\end{eqnarray}}
\def\ba{\begin{array}}
\def\ea{\end{array}}
\def\no{\nonumber}
\def\le{\langle}
\def\re{\rangle}
\def\lt{\left}
\def\rt{\right}

\newtheorem{Theorem}{Theorem}
\newtheorem{Definition}{Definition}
\newtheorem{Proposition}{Proposition}
\newtheorem{Lemma}{Lemma}
\newtheorem{Corollary}{Corollary}
\newcommand{\proof}[1]{{\bf Proof. }
        #1\begin{flushright}$\Box$\end{flushright}}

\baselineskip=20pt

\newfont{\elevenmib}{cmmib10 scaled\magstep1}
\newcommand{\preprint}{
   \begin{flushleft}
   \end{flushleft}\vspace{-1.3cm}
   \begin{flushright}\normalsize
   \end{flushright}}
\newcommand{\Title}[1]{{\baselineskip=26pt
   \begin{center} \Large \bf #1 \\ \ \\ \end{center}}}
\newcommand{\Author}{\begin{center}
   \large \bf
Junpeng Cao${}^{a}$,~Wen-Li Yang${}^b\footnote{Corresponding author:
wlyang@nwu.edu.cn}$,~Kangjie Shi${}^b$ and~Yupeng
Wang${}^a\footnote{Corresponding author: yupeng@iphy.ac.cn}$
 \end{center}}
\newcommand{\Address}{\begin{center}

     ${}^a$Beijing National Laboratory for Condensed Matter
           Physics, Institute of Physics, Chinese Academy of Sciences, Beijing
           100190, China\\
     ${}^b$Institute of Modern Physics, Northwest University,
     Xian 710069, China
   \end{center}}
\newcommand{\Accepted}[1]{\begin{center}
   {\large \sf #1}\\ \vspace{1mm}{\small \sf Accepted for Publication}
   \end{center}}

\preprint
\thispagestyle{empty}
\bigskip\bigskip\bigskip

\Title{Off-diagonal Bethe ansatz solutions of the anisotropic
spin-$\frac12$ chains with arbitrary boundary fields } \Author

\Address
\vspace{1cm}

\begin{abstract}
The anisotropic spin-$\frac{1}{2}$ chains with arbitrary boundary
fields are diagonalized with the off-diagonal Bethe ansatz method.
Based on the properties of the R-matrix and the K-matrices, an
operator product identity of the transfer matrix is constructed at
some special points of the spectral parameter. Combining with the
asymptotic behavior (for XXZ case) or the quasi-periodicity
properties (for XYZ case) of the transfer matrix,  the extended
$T-Q$ ansatzs and the corresponding Bethe ansatz equations are
derived.

\vspace{1truecm} \noindent {\it PACS:} 75.10.Pq, 03.65.Vf, 71.10.Pm

\end{abstract}
\newpage
\section{Introduction}
\label{intro} \setcounter{equation}{0}

The study of exactly solvable models (or quantum integrable systems)
\cite{Bax82} has attracted a great deal of interest  since Yang and
Baxter's pioneering works \cite{yang2,bax1}.  Such exact
non-perturbation  results have provided valuable insight into the
important universality classes of quantum physical systems ranging
from modern condensed matter physics \cite{Kas98} to string and
super-symmetric Yang-Mills theories \cite{Mad98,Dol03,Che04,Bei12}.
Moreover, quantum integrable models are paramount for the analysis
of nano-scale systems where alternative approaches involving mean
field approximations or perturbations have failed
\cite{Bla96,Duk04}.

The quantum inverse scattering method \cite{Kor93} (QISM) or the
algebraic Bethe ansatz method has been proven to be the most
powerful and (probably) unified method to construct exact solutions
to the spectrum problem of commuting families of conserved charges
(usually called the transfer matrix) in quantum integrable systems.
In the framework of QISM, the quantum Yang-Baxter equation (QYBE),
which defines the underlying algebraic structure,  has become a
cornerstone for constructing and solving the integrable models. So
far, there have been several well-known methods for deriving the
Bethe ansatz (BA) solutions of integrable models: the coordinate BA
\cite{Bet31,Alc87,Bel13,Cra12}, the T-Q approach \cite{Bax82,Yan06},
the algebraic BA \cite{Skl78,Tak79,Kor93,Fan96}, the analytic BA
\cite{Res83}, the functional BA \cite{Skl92} or the separation of
variables method among many others
\cite{And84,Baz89,Nep04,Cao03,Yan04,Gie05,Gie05-1,Mel05,Doi06,Baj06,Bas07,Gal08,Fra11,Nie09,Gra10,Nic12}.

Generally, there are two classes of integrable models: one possesses
$U(1)$ symmetry and the other does not. Three well-known examples
without $U(1)$ symmetry are the XYZ spin chain \cite{bax1,Tak79},
the spin chains with antiperiodic boundary condition
\cite{Yun95,Bat95,Gal08,Fra11,Nie09,Nic12} and the ones with
unparallel boundary fields
\cite{Nep04,Cao03,Yan04,Gie05,Bas07,Gal08,Fra11,Nie09,Gra10,Nic12}.
It has been proven that most of the conventional Bethe ansatz
methods can successfully diagonalize the integrable models with
$U(1)$ symmetry. However, for those without $U(1)$ symmetry, only
some very special cases such as the XYZ spin chain with even site
number \cite{bax1,Tak79} and the XXZ spin chain with constrained
unparallel boundary fields \cite{Fan96,Cao03,Yan04,Yan07} can be
dealt with due to the existence of a proper ``local vacuum state" in
these special cases. The main obstacle for applying the algebraic
Bethe ansatz and Baxter's method to general integrable models
without $U(1)$ symmetry lies in the absence of such a ``local
vacuum". A promising method for approaching such kind of problems is
Sklyanin's separation of variables method \cite{Skl92} which has
been recently applied to some integrable models
\cite{Fra11,Nie09,Gra10,Nic12}. However, before the very recent work
\cite{Cao13}, a systematic method was absent to derive the Bethe
ansatz equations for integrable models without $U(1)$ symmetry,
which are crucial for studying the physical properties in the
thermodynamic limit.

As for integrable models without $U(1)$ symmetry, the transfer
matrix contains not only the diagonal elements but also some
off-diagonal elements of the monodromy matrix. This breaks down the
usual $U(1)$ symmetry. Very recently, a systematic method
\cite{Cao13} for dealing with such kind of models was proposed by
the present authors, which has been shown \cite{Cao13-1} to
successfully construct the exact solutions of the open XXX chain
with unparallel boundary fields and the closed XYZ chain with odd
site number. The central idea of the method is to construct a proper
$T-Q$ ansatz with an extra off-diagonal term (comparing with the
ordinary ones) based on the functional relations between eigenvalues
$\Lambda(u)$ of the transfer matrix (the trace of the monodromy
matrix) and the quantum determinant $\Delta_q(u)$, (e.g. see below
(\ref{Eigen-Identity-6-1}) and (\ref{Eigen-Identity-8-1})) at some
special points of the spectral parameter $u=\theta_j$, i.e., \bea
\Lambda(\theta_j)\Lambda(\theta_j-\eta)\sim\Delta_q(\theta_j). \eea
Since the trace and the determinant are two basic quantities of a
matrix which are independent of the representation basis, this
method could overcome the obstacle of absence of a reference state
which is crucial in the conventional Bethe ansatz methods. Moreover,
in this paper we will show that the above relation can be lifted to
operator level (see below (\ref{Operator-id-3})) based on some
properties of the R-matrix and K-matrices.

Our primary motivation for this work comes from the
longstanding problem of solving the anisotropic spin-$\frac{1}{2}$
chain with arbitrary boundary fields, defined by the Hamiltonian
\cite{Skl88,Veg93}
\begin{eqnarray}
H = \sum_{j=1}^{N-1}
(J_x\sigma_j^x\sigma_{j+1}^x+J_y\sigma_j^y\sigma_{j+1}^y
+J_z\sigma_j^z\sigma_{j+1}^z)+ {\vec
h}^{(-)}\cdot\vec\sigma_{1}+\vec h^{(+)}\cdot\vec\sigma_{N},
\label{ohami}
\end{eqnarray}
where $N$ is the site number of the system; $\sigma_{j}^{\alpha}$
$(\alpha=x, y, z)$ is the Pauli matrix on the site $j$ along the
$\alpha$ direction; $\vec h^{(-)}=(h^{(-)}_x, h^{(-)}_y, h^{(-)}_z)$
and $\vec h^{(+)}=(h^{(+)}_x, h^{(+)}_y,h^{(+)}_z)$ are the boundary
magnetic fields; $J_\alpha$ $(\alpha=x, y, z)$ are the coupling
constants. Solving this problem for generic values of boundary
fields is a crucial step for understanding a variety of physical
systems without $U(1)$ symmetry. In this paper, we shall use the
method developed in \cite{Cao13} to solve the eigenvalue problem of
the above Hamiltonian with generic $\vec h^{(\pm)}$.

The paper is organized as follows.  Section 2 serves as an
introduction of our notations and some basic ingredients. We briefly
describe the inhomogeneous spin chains with periodic and
antiperiodic boundary conditions and open spin  chains with the most
general non-diagonal boundary terms.  In Section 3, based on the
properties of the $R$-matrix we derive the operator product
identities of the transfer matrices for the various closed and open
spin chains at some special points of the spectral parameter. In
section 4, the $T-Q$ ansatz for the eigenvalues of the transfer
matrix and the corresponding Bethe ansatz equations (BAEs) of the
open  XXZ spin chain are constructed based on the operator product
identities  of the transfer matrix and its  asymptotic behaviors.
Section 5 is attributed to the open XYZ spin chain. In section 6, we
summarize our results and give some discussions. Some detailed
technical proof is given in Appendix A$\&$B.


\section{ Transfer matrix}
\label{XXZ} \setcounter{equation}{0}

Throughout, ${\rm\bf V}$ denotes a two-dimensional linear space. The
R-matrix $R(u)\in {\rm End}({\rm\bf V}\otimes {\rm\bf V})$ is a
solution of the quantum Yang-Baxter equation (QYBE) \bea
R_{12}(u_1-u_2)R_{13}(u_1-u_3)R_{23}(u_2-u_3)
=R_{23}(u_2-u_3)R_{13}(u_1-u_3)R_{12}(u_1-u_2).\label{QYB}\eea It is
well-known that there are three standard classes of solutions (or
R-matrices), i.e., the elliptic (or the eight-vertex), the
trigonometric and rational (or six-vertex) R-matrix. The R-matrix is
given by \bea
R(u)=\lt(\begin{array}{llll}a(u)&&&d(u)\\&b(u)&c(u)&\\
&c(u)&b(u)&\\d(u)&&&a(u)\end{array}\rt). \label{r-matrix}
\eea The non-vanishing matrix elements of the eight-vertex R-matrix  are \cite{Bax82}
\bea
&&\hspace{-0.8truecm}a(u)\hspace{-0.1truecm}=
\hspace{-0.1truecm}\frac{\theta\lt[\begin{array}{c} 0\\\frac{1}{2}
\end{array}\rt]\hspace{-0.16truecm}(u,2\tau)\hspace{0.12truecm}
\theta\lt[\begin{array}{c} \frac{1}{2}\\[2pt]\frac{1}{2}
\end{array}\rt]\hspace{-0.16truecm}(u+\eta,2\tau)}{\theta\lt[\begin{array}{c} 0\\\frac{1}{2}
\end{array}\rt]\hspace{-0.16truecm}(0,2\tau)\hspace{0.12truecm}
\theta\lt[\begin{array}{c} \frac{1}{2}\\[2pt]\frac{1}{2}
\end{array}\rt]\hspace{-0.16truecm}(\eta,2\tau)},\quad
b(u)\hspace{-0.1truecm}=\hspace{-0.1truecm}\frac{\theta\lt[\begin{array}{c}
\frac{1}{2}\\[2pt]\frac{1}{2}
\end{array}\rt]\hspace{-0.16truecm}(u,2\tau)\hspace{0.12truecm}
 \theta\lt[\begin{array}{c} 0\\\frac{1}{2}
\end{array}\rt]\hspace{-0.16truecm}(u+\eta,2\tau)}
{\theta\lt[\begin{array}{c} 0\\\frac{1}{2}
\end{array}\rt]\hspace{-0.16truecm}(0,2\tau)\hspace{0.12truecm}
\theta\lt[\begin{array}{c} \frac{1}{2}\\[2pt]\frac{1}{2}
\end{array}\rt]\hspace{-0.16truecm}(\eta,2\tau)},\label{Eight-1}\\
&&\hspace{-0.8truecm}c(u)\hspace{-0.1truecm}=
\hspace{-0.1truecm}\frac{\theta\lt[\begin{array}{c} 0\\\frac{1}{2}
\end{array}\rt]\hspace{-0.16truecm}(u,2\tau)\hspace{0.12truecm}
 \theta\lt[\begin{array}{c} 0\\\frac{1}{2}
\end{array}\rt]\hspace{-0.16truecm}(u+\eta,2\tau)}
{\theta\lt[\begin{array}{c} 0\\\frac{1}{2}
\end{array}\rt]\hspace{-0.16truecm}(0,2\tau)\hspace{0.12truecm}
\theta\lt[\begin{array}{c} 0\\\frac{1}{2}
\end{array}\rt]\hspace{-0.16truecm}(\eta,2\tau)},\quad
d(u)\hspace{-0.1truecm}=\hspace{-0.1truecm}\frac{\theta\lt[\begin{array}{c}
\frac{1}{2}\\[2pt]\frac{1}{2}
\end{array}\rt]\hspace{-0.16truecm}(u,2\tau)\hspace{0.12truecm}
 \theta\lt[\begin{array}{c} \frac{1}{2}\\[2pt]\frac{1}{2}
\end{array}\rt]\hspace{-0.16truecm}(u+\eta,2\tau)}
{\theta\lt[\begin{array}{c} 0\\\frac{1}{2}
\end{array}\rt]\hspace{-0.16truecm}(0,2\tau)\hspace{0.12truecm}
\theta\lt[\begin{array}{c} 0\\\frac{1}{2}
\end{array}\rt]\hspace{-0.16truecm}(\eta,2\tau)},\label{Eight-2}
\eea while the non-vanishing matrix elements of the six-vertex R-matrix
are \cite{Kor93}
\bea
a(u)&=&\frac{\varphi(u+\eta)}{\varphi(\eta)},\quad b(u)=\frac{\varphi(u)}{\varphi(\eta)},\label{Six-1}\\
c(u)&=&1,\quad d(u)=0.\label{Six-2}
\eea Here the generic complex number $\eta$ is the so-called crossing parameter, the definitions of elliptic functions are given in Appendix A
and the function $\varphi(u)$ is defined as
\bea
\varphi(u)=\left\{\begin{array}{ll}\sinh(u)&{\rm for\,the\,trigonometric\,case},\\
u&{\rm for\,the\,rational\,case}.\end{array}\rt.\no \eea All the
R-matrices possess the following   properties, \bea
&&\hspace{-1.5cm}\mbox{ Initial
condition}:\,R_{12}(0)= P_{12},\label{Int-R}\\
&&\hspace{-1.5cm}\mbox{ Unitarity
relation}:\,R_{12}(u)R_{21}(-u)= -\xi(u)\,{\rm id},\label{Unitarity}\\
&&\hspace{-1.5cm}\mbox{ Crossing
relation}:\,R_{12}(u)=V_1R_{12}^{t_2}(-u-\eta)V_1,\quad
V=-i\s^y,
\label{crosing-unitarity}\\
&&\hspace{-1.5cm}\mbox{ PT-symmetry}:\,R_{12}(u)=R_{21}(u)=R^{t_1\,t_2}_{12}(u),\label{PT}\\
&&\hspace{-1.5cm}\mbox{$Z_2$-symmetry}: \,
\qquad\s^i_1\s^i_2R_{1,2}(u)=R_{1,2}(u)\s^i_1\s^i_2,\quad
\mbox{for}\,\,
i=x,y,z,\label{Z2-sym}\\
&&\hspace{-1.5cm}\mbox{ Antisymmetry}:\,R_{12}(-\eta)=-(1-P)=-2P^{(-)}.\label{Ant}
\eea
Here $\s^i$ ($i=x,\,y,\,z$) is the usual Pauli matrix,  $R_{21}(u)=P_{12}R_{12}(u)P_{12}$ with $P_{12}$ being
the usual permutation operator and $t_i$ denotes transposition
in the $i$-th space. The function $\xi(u)$ is given by
\bea
\xi(u)=\left\{\begin{array}{ll}\frac{\s(u+\eta)\s(u-\eta)}{\s(\eta)\s(\eta)}
&{\rm for\,\,the\,\,eight-vertex\,\,case},\\
\frac{\varphi(u+\eta)\varphi(u-\eta)}{\varphi(\eta)\varphi(\eta)}
&{\rm for\,\,the\,\,six-vertex\,\,case}.\end{array}\rt.\label{xi-function}
\eea Here and below we adopt the standard
notations: for any matrix $A\in {\rm End}({\rm\bf V})$, $A_j$ is an
embedding operator in the tensor space ${\rm\bf V}\otimes
{\rm\bf V}\otimes\cdots$, which acts as $A$ on the $j$-th space and as
identity on the other factor spaces; $R_{ij}(u)$ is an embedding
operator of R-matrix in the tensor space, which acts as identity
on the factor spaces except for the $i$-th and $j$-th ones.

We introduce the ``row-to-row"  (or one-row ) monodromy matrices
$T_0(u)$ and $\hat{T}_0(u)$, which are $2\times 2$ matrices with
elements being operators acting on ${\rm\bf V}^{\otimes N}$, \bea
T_0(u)&=&R_{0N}(u-\theta_N)R_{0\,N-1}(u-\theta_{N-1})\cdots
R_{01}(u-\theta_1),\label{Mon-V-1}\\
\hat{T}_0(u)&=&R_{01}(u+\theta_1)R_{02}(u+\theta_{2})\cdots
R_{0N}(u+\theta_N).\label{Mon-V-2} \eea Here
$\{\theta_j|j=1,\cdots,N\}$ are arbitrary free complex parameters
which are usually called as inhomogeneous parameters. The transfer
matrix $t^{(p)}$ of the spin chain with periodic boundary condition
(or closed chain) is given by \cite{Kor93} \bea
t^{(p)}(u)=tr_0(T_0(u)).\label{trans-Periodic} \eea The QYBE
(\ref{QYB}) leads to the fact that the transfer matrices with
different spectral parameters commute with each other \cite{Skl78}:
$[t^{(p)}(u),t^{(p)}(v)]=0$. Then $t^{(p)}(u)$ serves as the
generating functional of the conserved quantities of the
corresponding system, which ensures the integrability of the closed
spin chain.

Integrable open chain can be constructed as follows \cite{Skl88}.
Let us introduce a pair of K-matrices $K^-(u)$ and $K^+(u)$. The
former satisfies the reflection equation (RE)
 \bea &&R_{12}(u_1-u_2)K^-_1(u_1)R_{21}(u_1+u_2)K^-_2(u_2)\no\\
 &&~~~~~~=
K^-_2(u_2)R_{12}(u_1+u_2)K^-_1(u_1)R_{21}(u_1-u_2),\label{RE-V}\eea
and the latter  satisfies the dual RE \bea
&&R_{12}(u_2-u_1)K^+_1(u_1)R_{21}(-u_1-u_2-2)K^+_2(u_2)\no\\
&&~~~~~~= K^+_2(u_2)R_{12}(-u_1-u_2-2)K^+_1(u_1)R_{21}(u_2-u_1).
\label{DRE-V}\eea For open spin chains, other than the standard
``row-to-row" monodromy matrix $T_0(u)$ (\ref{Mon-V-1}), one needs
to consider  the
 double-row monodromy matrix $\mathbb{T}_0(u)$
\bea
  \mathbb{T}_0(u)=T_0(u)K^-_0(u)\hat{T}_0(u).
  \label{Mon-V-0}
\eea Then the double-row transfer matrix $t(u)$ of the spin chain
with open boundary (or the open spin chain) is given by \bea
t(u)=tr_0(K^+_0(u)\mathbb{T}_0(u)).\label{trans}\eea The QYBE
(\ref{QYB}) and (dual) REs (\ref{RE-V}) and (\ref{DRE-V}) lead to
the fact that the transfer matrices with different spectral
parameters commute with each other \cite{Skl88}: $[t(u),t(v)]=0$.
Then $t(u)$ serves as the generating functional of the conserved
quantities of the corresponding system, which ensures the
integrability of the open spin chain.

The integrable Hamiltonian (\ref{ohami}) can be obtained from the
transfer matrix as follows. For the XXZ case
\begin{eqnarray}
&&H=\sinh\eta \frac{\partial \ln t(u)}{\partial
u}|_{u=0,\theta_j=0}-N\cosh\eta -\tanh\eta\sinh\eta \nonumber \\
&&\quad =2\sinh\eta\sum_{j=1}^{N-1}P_{j,j+1}R_{j,j+1}^\prime(0) +\sinh\eta
\frac{tr_0 {K^+_0}^\prime(0)}{tr_0 K^+_0(0)} +2\sinh\eta\frac{tr_0 K_{0}^+
P_{N,0}R_{N,0}^\prime(0)}{tr_0 K^+_0(0)}
\nonumber\\
&&\qquad\qquad +\sinh\eta \frac{{K_{1}^-}^\prime(0)}{K^-_1(0)}-N\cosh\eta -\tanh\eta\sinh\eta\nonumber\\
&&\quad
=\sum_{j=1}^{N-1}\left[\sigma_{j}^{x}\sigma_{j+1}^{x}+\sigma_{j}^{y}
\sigma_{j+1}^{y}+\cosh\eta\sigma_{j}^{z} \sigma_{j+1}^{z}\right]
\nonumber \\
&&\qquad +\frac{ \sinh\eta}{\sinh\alpha_-\cosh\beta_-}
(\cosh\alpha_-\sinh\beta_- \sigma_{1}^z + \cosh\theta_-\sigma_1^x +
i\sinh\theta_- \sigma_1^y) \nonumber \\
&&\qquad + \frac{ \sinh\eta}{\sinh\alpha_+\cosh\beta_+}
(-\cosh\alpha_+\sinh\beta_+ \sigma_{N}^z + \cosh\theta_+\sigma_N^x +
i\sinh\theta_+ \sigma_N^y). \label{oh}
\end{eqnarray} The corresponding K-matrices are the most general solutions $K^{\mp}(u)$ in \cite{Veg93,Gho94} (see below the
Eqs.(\ref{K-matrix})-(\ref{K-6-2})). For the XYZ case, \bea
H&=&\frac{\s(\eta)}{\s'(0)}\lt\{\lt.\frac{\partial}{\partial u}\,\ln
t(u)\rt|_{u=0}-\lt[(N-1)\zeta(\eta)+2\zeta(2\eta)\rt]\rt\}\nonumber\\
&=&\sum^{N-1}_{j=1}\lt(J_x\s^x_j\s^x_{j+1}
+J_y\s^y_j\s^y_{j+1}+J_z\s^z_j\s^z_{j+1}\rt)\,+\, h^{(-)}_x\s^x_1
+h^{(-)}_y\s^y_1+h^{(-)}_z\s^z_1\no\\
&&\qquad + h^{(+)}_x\s^x_N
+h^{(+)}_y\s^y_N+h^{(+)}_z\s^z_N,\label{Ham}\eea with \bea
J_x=\frac{e^{i\pi\eta}\s(\eta+\frac{\tau}{2})}{\s(\frac{\tau}{2})},
\quad J_y=\frac{e^{i\pi\eta}\s(\eta+\frac{1}{2}+\frac{\tau}{2})}
{\s(\frac{1}{2}+\frac{\tau}{2})}, \quad
J_z=\frac{\s(\eta+\frac{1}{2})}{\s(\frac{1}{2})}, \no\eea and  \bea
h^{(\mp)}_z&=&\pm \frac{\s(\eta)}
{\s(\frac{1}{2})}\prod_{l=1}^3\frac{\s(\a^{(\mp)}_l-\frac{1}{2})}
{\s(\a^{(\mp)}_l)},\no\\ h^{(\mp)}_x&=&\pm
e^{-i\pi(\sum_{l=1}^3\a^{(\mp)}_l-\frac{\tau}{2})}\frac{\s(\eta)}
{\s(\frac{\tau}{2})}\prod_{l=1}^3\frac{\s(\a^{(\mp)}_l-\frac{\tau}{2})}
{\s(\a^{(\mp)}_l)},\no \\
h^{(\mp)}_y&=&\pm
e^{-i\pi(\sum_{l=1}^3\a^{(\mp)}_l-\frac{1}{2}-\frac{\tau}{2})}\frac{\s(\eta)}
{\s(\frac{1}{2}+\frac{\tau}{2})}\prod_{l=1}^3
\frac{\s(\a^{(\mp)}_l-\frac{1}{2}-\frac{\tau}{2})}
{\s(\a^{(\mp)}_l)}. \label{Boundary}\eea Here $\s(u)$ is the
$\s$-function defined by (\ref{Function}) and $\{\a^{(\mp)}_l\}$ are
parameters contained in the most general K-matrices
\cite{Ina94,Hou95} (see below Eqs.(\ref{K-matrix1}) and
(\ref{K-matrix2})).


\section{ Operator identity of the transfer matrix}
\label{T-QR} \setcounter{equation}{0}

\subsection{Closed chains}

In order to get some functional relations of the transfer matrix, we
evaluate the transfer matrix $t^{(p)}(u)$ at some particular points
such as $\theta_j$ and $\theta_j-\eta$. Using the similar procedure as in
\cite{Kit99,Mai00}, we apply the initial condition
(\ref{Int-R}) of the R-matrix to express the transfer
matrix $t^{(p)}(\theta_j)$ as \bea
t^{(p)}(\theta_j)&=&tr_{0}\lt\{R_{0N}(\theta_j-\theta_N)\ldots
R_{0\,j+1}(\theta_j-\theta_{j+1})P_{0j}R_{0\,j-1}(\theta_j-\theta_{j-1})\ldots
 R_{01}(\theta_j-\theta_1)\rt\}\no\\
&=&R_{j\,j-1}(\theta_j-\theta_{j-1})\ldots R_{j1}(\theta_j-\theta_1)
tr_{0}\lt\{R_{0N}(\theta_j-\theta_N)\ldots R_{0\,j+1}(\theta_j-\theta_{j+1})P_{0j}\rt\}\no\\
&=&R_{j\,j-1}(\theta_j-\theta_{j-1})\ldots R_{j1}(\theta_j-\theta_1)R_{jN}(\theta_j-\theta_N)\ldots
R_{j\,j+1}(\theta_j-\theta_{j+1}).\no
\eea In deriving the above equation, we have used the identity: $tr_0(P_{0j})={\rm id}_j$. The crossing relation (\ref{crosing-unitarity}) and $V^2=-1$ enable one to express the transfer matrix $t^{(p)}(\theta_j-\eta)$ as
\bea
t^{(p)}(\theta_j-\eta)&=&(-1)^N\,tr_0\lt\{ R^{t_0}_{0N}(-\theta_j+\theta_N)\ldots R^{t_0}_{01}(-\theta_j+\theta_1)\rt\}\no\\
&=&(-1)^N\,tr_0\lt\{ R_{01}(-\theta_j+\theta_1)\ldots R_{0N}(-\theta_j+\theta_N)\rt\}\no\\
&=&(-1)^{N}R_{j\,j+1}(-\theta_j+\theta_{j+1})\ldots R_{jN}(-\theta_j+\theta_{N})\no\\
&&\times R_{j1}(-\theta_j+\theta_1)\ldots
R_{j\,j-1}(-\theta_j+\theta_{j-1}).\no
\eea Using the unitarity relation (\ref{Unitarity}), we have
\bea
 t^{(p)}(\theta_j) t^{(p)}(\theta_j-\eta)=\Delta^{(p)}_q(\theta_j)=\prod_{l=1}^N\xi(\theta_j-\theta_l)\times {\rm id},\label{operator-id-1}
\eea where the function $\xi(u)$ is given by (\ref{xi-function}).
The commutativity of the transfer matrices with different spectrum
implies that they have  common eigenstates. Let $|\Psi\rangle$ be an
eigenstate of $t^{(p)}(u)$, which does not depend upon $u$,  with
the eigenvalue $\Lambda^{(p)}(u)$, i.e., \bea
t^{(p)}(u)|\Psi\rangle=\Lambda^{(p)}(u)|\Psi\rangle.\no \eea The
very operator identity (\ref{operator-id-1}) leads to that the
corresponding eigenvalue $\Lambda^{(p)}(u)$ satisfies the following
relation \bea \Lambda^{(p)}(\theta_j)
\Lambda^{(p)}(\theta_j-\eta)=\prod_{l=1}^N\xi(\theta_j-\theta_l),\quad
j=1,\ldots,N.\label{Eigen-id-1} \eea

Some remarks are in order. The QYBE (\ref{QYB}) and the $Z_2$-symmetry (\ref{Z2-sym}) of the R-matrix imply that the corresponding spin chain with
antiperiodic (or twisted ) boundary condition is also integrable, the corresponding transfer matrix $t^{(ap)}(u)$ can be given by
\bea
 t^{(ap)}(u)=tr_0(\sigma_0^x T_0(u)).\label{trans-AntiPeriodic}
\eea Noticing that $V\sigma^x=-\sigma^x V$,  one can easily
derive the following relation \bea
  t^{(ap)}(\theta_j)t^{(ap)}(\theta_j-\eta)=-\Delta^{(p)}_q(\theta_j)=-\prod_{l=1}^N\xi(\theta_j-\theta_l)\times {\rm id}.\label{operator-id-2}
\eea This leads to that the corresponding eigenvalue $\Lambda^{(ap)}(u)$ satisfies the following relation
\bea
\Lambda^{(ap)}(\theta_j) \Lambda^{(ap)}(\theta_j-\eta)=-\prod_{l=1}^N\xi(\theta_j-\theta_l),\quad j=1,\ldots,N.\label{Eigen-id-2}
\eea The above relations for the trigonometric case were obtained in \cite{Cao13} by solving some recursion relation.
Similar relations were also derived in \cite{Nic12} by the separation of variables method.

\subsection{Open chains}
Now we are in position to compute the transfer matrix $t(u)$ of the open chain at $\theta_j$ and $\theta_j-\eta$. Following the similar
procedure as in \cite{Hik95, Wan00}, we have  that
\bea
 t(\theta_j)&=&R_{j\,j-1}(\theta_j-\theta_{j-1})\ldots R_{j1}(\theta_j-\theta_1)K^-_j(\theta_j)R_{1j}(\theta_1+\theta_j)\ldots R_{j-1\,j}(\theta_{j-1}+\theta_j)\no\\
&&\times tr_0\lt\{ K_0^+(\theta_j)R_{0N}(\theta_j-\theta_N)\ldots R_{0\,j+2}(\theta_j-\theta_{j+2})R_{0\,j+1}(\theta_j-\theta_{j+1})\rt.\no\\
&&\quad\times\lt.P_{0j}R_{j0}(2\theta_j)R_{j+1\,0}(\theta_{j+1}+\theta_j) R_{j+2\,0}(\theta_{j+2}+\theta_j)\ldots R_{N\,0}(\theta_{N}+\theta_j)
\rt\}\no
\eea Using QYBE (\ref{QYB}), we have
\bea
&&R_{0\,j+1}(\theta_j-\theta_{j+1})P_{0j}R_{j0}(2\theta_j)R_{j+1\,0}(\theta_{j+1}+\theta_j)\no\\
&&\quad\quad =
R_{0\,j+1}(\theta_j-\theta_{j+1})R_{0j}(2\theta_j)R_{j+1\,j}(\theta_{j+1}+\theta_j)P_{0j}\no\\
&&\quad \quad =R_{j+1\,j}(\theta_{j+1}+\theta_{j})R_{0j}(2\theta_j)R_{0\,j+1}(\theta_{j}-\theta_{j+1})P_{0j}\no\\
&&\quad\quad =R_{j+1\,j}(\theta_{j+1}+\theta_{j})P_{0j}R_{j0}(2\theta_j)R_{j\,j+1}(\theta_{j}-\theta_{j+1}).\no
\eea This gives rise to
\bea
 t(\theta_j)&=&R_{j\,j-1}(\theta_j-\theta_{j-1})\ldots R_{j1}(\theta_j-\theta_1)K^-_j(\theta_j)
R_{1j}(\theta_1+\theta_j)\ldots R_{j-1\,j}(\theta_{j-1}+\theta_j)\no\\
&&\times R_{j+1\,j}(\theta_{j+1}+\theta_j)\ldots R_{Nj}(\theta_{N}+\theta_j)
tr_0\{K^+_0(\theta_j)P_{0j}R_{j0}(2\theta_j)\}\no\\
&&\times R_{jN}(\theta_{j}-\theta_N)\ldots R_{j\,j+1}(\theta_{j}-\theta_{j+1}).\label{t-theta-j}
\eea The crossing relation (\ref{crosing-unitarity}) of the R-matrix implies
\bea
\hspace{-1.2truecm} t(\theta_j-\eta)&=&tr_0\lt\{V_0K^+_0(\theta_j-\eta)V_0R^{t_0}_{0N}(-\theta_j+\theta_N)\ldots
R^{t_0}_{01}(-\theta_j+\theta_1)\rt.\no\\
&&\quad\times \lt.V_0K_0^-(\theta_j-\eta)V_0 R^{t_0}_{10}(-\theta_1-\theta_j)
\ldots R^{t_0}_{0N}(-\theta_N-\theta_j)
\rt\} \no\\
&=&tr_0\lt\{\lt(V_0K^+_0(\theta_j-\eta)V_0R^{t_0}_{0N}(-\theta_j+\theta_N)\ldots
R^{t_0}_{01}(-\theta_j+\theta_1)\rt)^{t_0}\rt.\no\\
&&\quad\times \lt.(V_0K_0^-(\theta_j-\eta)V_0 R^{t_0}_{10}(-\theta_1-\theta_j)
\ldots R^{t_0}_{0N}(-\theta_N-\theta_j))^{t_0}
\rt\} \no\\
&=&tr_0\lt\{\lt(V_0K^-_0(\theta_j-\eta)V_0\rt)^{t_0}R_{01}(-\theta_j+\theta_1)\ldots
R_{0N}(-\theta_j+\theta_N)\rt.\no\\
&&\quad\times \lt.\lt(V_0K_0^+(\theta_j-\eta)V_0\rt)^{t_0} R_{N0}(-\theta_N-\theta_j)
\ldots R_{10}(-\theta_1-\theta_j)\rt\} \no\\
&=& R_{j\,j+1}(-\theta_j+\theta_{j+1})\ldots R_{jN}(-\theta_j+\theta_{N})
\lt\{V_jK^+_j(\theta_j-\eta)V_j\rt\}^{t_j}\no\\
&&\times R_{Nj}(\hspace{-0.08truecm}-\hspace{-0.08truecm}\theta_N\hspace{-0.08truecm}-\hspace{-0.08truecm}\theta_{j})
\ldots R_{j+1\,j}(\hspace{-0.08truecm}-\hspace{-0.08truecm}\theta_{j+1}\hspace{-0.08truecm}-\hspace{-0.08truecm}\theta_{j})
R_{j-1\,j}(\hspace{-0.08truecm}-\hspace{-0.08truecm}\theta_{j-1}\hspace{-0.08truecm}-\hspace{-0.08truecm}\theta_{j})\ldots
R_{1j}(\hspace{-0.08truecm}-\hspace{-0.08truecm}\theta_{1}\hspace{-0.08truecm}-\hspace{-0.08truecm}\theta_{j})\no\\
&&\times tr_0\lt\{
\lt(V_0K^-_0(\theta_j-\eta)V_0 \rt)^{t_0}P_{0j} R_{j0}(-2\theta_j)\rt\}\no\\
&&\times R_{j1}(-\theta_{j}+\theta_{1})\ldots
R_{j\,j-1}(-\theta_{j}+\theta_{j-1}). \eea With the help of the
unitary relation (\ref{Unitarity}) of the R-matrix, we find that the
transfer matrix satisfies the following relations \bea t(\theta_j)
t(\theta_j-\eta)=-\frac{\Delta^{(o)}_q(\theta_j)}{\xi(2\theta_j)}.
\label{Operator-id-3} \eea For generic $\{\theta_j\}$, the quantum
determinant operator is proportional to the identity operator \bea
 \Delta^{(o)}_q(u)=\delta(u)\times {\rm id},
\eea The expression of the function $\delta(u)$ is given by \cite{Mez90,Zho96}
\bea
\delta(u)={\rm Det}_q\{T(u)\}\, {\rm Det}_q\{\hat{T}(u)\}
\,{\rm Det}_q\{K^-(u)\}\,{\rm Det}_q\{K^+(u)\},\label{Det-function-1}
\eea and the above various determinants are
 \bea
\rm{Det}_q\lt\{T(u)\rt\}\,{\rm
id}&=&tr_{12}\lt(P^{(-)}_{12}T_1(u-\eta)T_2(u)P^{(-)}_{12}\rt)
=\prod_{j=1}^N\xi(u-\theta_j)\,{\rm
id},\no\\
\rm{Det}_q\lt\{\hat{T}(u)\rt\}\,{\rm
id}&=&tr_{12}\lt(P^{(-)}_{12}\hat{T}_1(u-\eta)\hat{T}_2(u)P^{(-)}_{12}\rt)
=\prod_{j=1}^N\xi(u+\theta_j)\,{\rm
id},\no\\
\rm{Det}_q\lt\{K^-(u)\rt\}&=&tr_{12}
\lt(P^{(-)}_{12}K^-_1(u-\eta)R_{12}(2u-\eta)K^-_2(u)\rt),\label{Det-function-2}\\
\rm{Det}_q\lt\{K^+(u)\rt\}&=&tr_{12}
\lt(P^{(-)}_{12}K^+_2(u)R_{12}(-2u-\eta)K^+_1(u-\eta)\rt).\label{Det-function-3}
\eea In the derivation of the functional relation
(\ref{Operator-id-3}), the following identities have been  used
\bea
K^-_j(u)\,tr_0\lt\{(V_0K^-_0(u-\eta)V_0)^{t_0}P_{0j}R_{j0}(-2u)\rt\}&=&
{\rm Det}_q\lt\{K^-(u)\rt\}\times {\rm id}_j, \label{Id-function-1}\\
tr_0\lt\{K^+_0(u)P_{0j}R_{j0}(2u)\rt\}\,\lt\{V_jK^+_j(u-\eta)V_j\rt\}^{t_j}&=&
{\rm Det}_q\lt\{K^+(u)\rt\}\times {\rm id}_j. \label{Id-function-2}
\eea The proof of the above  equations is relegated to Appendix B.

In the following two sections, we shall demonstrate how to use the
above operator identity to construct the off-diagonal Bethe ansatz
solutions of the open chains with the most general boundary terms.


\section{ Results for the open XXZ chain}
\label{BAE} \setcounter{equation}{0}
\subsection{Functional relations}
The most general solutions $K^{\mp}(u)$ \cite{Veg93,Gho94} to the
reflection equation and its dual associated with the trigonometric
six-vertex R-matrix (for the rational case see \cite{Cao13-1}), or
the XXZ chain, are given respectively by \bea
K^-(u)&=&\lt(\begin{array}{ll}K^-_{11}(u)&K^-_{12}(u)\\
K^-_{21}(u)&K^-_{22}(u)\end{array}\rt),\no\\
K^-_{11}(u)&=&2\lt(\sinh(\a_-)\cosh(\b_{-})\cosh(u)
+\cosh(\a_-)\sinh(\b_-)\sinh(u)\rt),\no\\
K^-_{22}(u)&=&2\lt(\sinh(\a_-)\cosh(\b_{-})\cosh(u)
-\cosh(\a_-)\sinh(\b_-)\sinh(u)\rt),\no\\
K^-_{12}(u)&=&e^{\theta_-}\sinh(2u),\quad
K^-_{21}(u)=e^{-\theta_-}\sinh(2u),\label{K-matrix}\eea and \bea
K^+(u)=\lt.K^-(-u-\eta)\rt|_{(\a_-,\b_-,\theta_-)\rightarrow
(-\a_+,-\b_+,\theta_+)}.\label{K-6-2} \eea Here
$\a_{\mp},\,\b_{\mp},\,\theta_{\mp}$ are the boundary parameters
which are associated with boundary field terms (see (\ref{oh})). Then the associated
function $\delta(u)$ defined by (\ref{Det-function-1}) reads
\cite{Nep04} \bea \d(u)&=&-2^4\frac{\sinh(2u-2\eta)
\sinh(2u+2\eta)}{\sinh^2\eta}\sinh(u+\a_-)\sinh(u-\a_-)
\cosh(u+\b_-)\cosh(u-\b_-)\no\\
&&\quad\times \sinh(u+\a_+)\sinh(u-\a_+)
\cosh(u+\b_+)\cosh(u-\b_+)\no\\
&&\quad\times
\prod_{l=1}^N\frac{\sinh(u+\theta_l+\eta)\sinh(u-\theta_l+\eta)\sinh(u+\theta_l-\eta)\sinh(u-\theta_l-\eta)}
{\sinh^4(\eta)}. \label{delta-function-trigono} \eea Following the
method in \cite{Yan08-e,Beh96} and using the crossing relation of the
R-matrix (\ref{crosing-unitarity}) and the explicit expressions of
the K-matrices (\ref{K-matrix}) and (\ref{K-6-2}), one can show that
the corresponding transfer matrix $t(u)$ satisfies the following
crossing relation \bea t(-u-\eta)&=&t(u).\label{crosing-opertaor-6}
\eea The quasi-periodicity of the R-matrix and K-matrices \bea
R_{12}(u+i\pi)=-\s^z_1\,R_{12}(u)\,\s^z_1=-\s^z_2\,R_{12}(u)\,\s^z_2,\quad
K^{\pm}(u+i\pi)=-\s^z\,K^{\pm}(u)\,\s^z,\no \eea and the explicit
expression of the K-matrix $K^-(u)$ given by (\ref{K-matrix}) and
its special values  at $u=0,\,\frac{i\pi}{2}$: \bea
K^-(0)=\frac{1}{2}tr(K^-(0))\times {\rm id},\quad
K^-(\frac{i\pi}{2})=\frac{1}{2}tr(K^-(\frac{i\pi}{2})\s^z)\times
\s^z,\no \eea allow one to derive   that the associated transfer
matrix satisfies the following properties \bea
t(u+i\pi)&=&t(u),\label{Periodic-6}\\
t(0)&=&-2^3\sinh\a_-\cosh\b_-\sinh\a_+\cosh\b_+\cosh\eta\,\no\\
&&\times
\prod_{l=1}^N\frac{\sinh(\eta-\theta_l)\,\sinh(\eta+\theta_l)}{\sinh^2\eta}\times
{\rm id},\\
t(\frac{i\pi}{2})&=&-2^3\cosh\a_-\sinh\b_-\cosh\a_+\sinh\b_+\cosh\eta\,\no\\
&&\times
\prod_{l=1}^N\frac{\sinh(\frac{i\pi}{2}+\theta_l+\eta)\sinh(\frac{i\pi}{2}+\theta_l-\eta)}{\sinh^2\eta}
\times {\rm id},\\
\lim_{u\rightarrow \pm\infty}
t(u)&=&-\frac{\cosh(\theta_--\theta_+)e^{\pm[(2N+4)u+(N+2)\eta]}}
{2^{2N+1}\sinh^{2N}\eta}\times {\rm id} +\ldots.\label{Tran-Asy-6}
\eea We shall demonstrate how the very identity
(\ref{Operator-id-3}), the explicit expression
(\ref{delta-function-trigono}) of the model, and
(\ref{crosing-opertaor-6})-(\ref{Tran-Asy-6}) allow us to completely
determine the eigenvalues of the corresponding transfer matrix of
the open XXZ chain with the most generic  K-matrices given by
(\ref{K-matrix})-(\ref{K-6-2}).

The commutativity of the transfer matrix $t(u)$ implies that one can
find the common eigenstates of $t(u)$, which indeed do not depend
upon $u$. Suppose $|\Psi\rangle$ is an eigenstate of $t(u)$ with an
eigenvalue $\Lambda(u)$, namely, \bea t(u)|\Psi\rangle
=\Lambda(u)|\Psi\rangle.\no \eea The very operator identity
(\ref{Operator-id-3}) of the six-vertex model implies the
corresponding eigenvalue $\Lambda(u)$  satisfies the similar
relation \bea
\Lambda(\theta_j)\Lambda(\theta_j-\eta)=\frac{\delta(\theta_j)\,\sinh\eta\,\sinh\eta}{\sinh(\eta-2\theta_j)\,\sinh(\eta+2\theta_j)},
\quad j=1,\ldots,N, \label{Eigen-Identity-6-1} \eea where the
function $\delta(u)$ is given by (\ref{delta-function-trigono}).
Similar relation for the open XXX chain was also previously obtained
in \cite{Fra08} by the separation of variables method. The special case of the identity (\ref{Eigen-Identity-6-1}),
when one of K-matrices $K^{\pm}(u)$ is  diagonal or lower triangle matrix,  was  derived in \cite{Nic13} by the separation of variables method.

The
properties of the transfer matrix $t(u)$ given by
(\ref{crosing-opertaor-6})-(\ref{Tran-Asy-6}) imply that the
corresponding eigenvalue $\Lambda(u)$ satisfies the following
relations: \bea
\L(-u-\eta)&=&\L(u),\quad \L(u+i\pi)=\L(u),\label{crosing-Eign-6}\\
\L(0)&=&-2^3\sinh\a_-\cosh\b_-\sinh\a_+\cosh\b_+\cosh\eta\,\no\\
&&\times \prod_{l=1}^N\frac{\sinh(\eta-\theta_l)\,\sinh(\eta+\theta_l)}{\sinh^2\eta},\label{Eigen-6-1}\\
\L(\frac{i\pi}{2})&=&-2^3\cosh\a_-\sinh\b_-\cosh\a_+\sinh\b_+\cosh\eta\,\no\\
&&\times \prod_{l=1}^N\frac{\sinh(\frac{i\pi}{2}+\theta_l+\eta)\,\sinh(\frac{i\pi}{2}+\theta_l-\eta)}{\sinh^2\eta},\\
\lim_{u\rightarrow \pm\infty}
\L(u)&=&-\frac{\cosh(\theta_--\theta_+)e^{\pm[(2N+4)u+(N+2)\eta]}}
{2^{2N+1}\sinh^{2N}\eta}+\ldots.\label{Eigen-Asy-6}
\eea The asymptotic behavior (\ref{Eigen-Asy-6}), the second
relation of (\ref{crosing-Eign-6}), and the analyticity of the
R-matrix and K-matrices and independence on $u$ of the eigenstate
lead to the fact  that the eigenvalue $\Lambda(u)$ further possesses
the property \bea \L(u) \mbox{, as an entire function of $u$, is a
trigonometric polynomial of degree $2N+4$}.\label{Eigen-Anal-6} \eea
Therefore  the very functional relations
(\ref{Eigen-Identity-6-1})-(\ref{Eigen-Anal-6}) completely determine
the function $\L(u)$. Here we construct the solutions of these
equations in terms of a generalized  $T-Q$ ansatz formulism
developed in \cite{Cao13}. For this purpose, we introduce the
following functions \bea
\bar{A}(u)&=&\prod_{l=1}^N\frac{\sinh(u-\theta_l+\eta)\,\sinh(u+\theta_l+\eta)}{\sinh^2\eta},\no\\
\bar{a}(u)&=&-2^2\frac{\sinh(2u+2\eta)}{\sinh(2u+\eta)}\sinh(u-\a_-)\cosh(u-\b_-)\no\\
&&\quad\quad\times\sinh(u-\a_+)\cosh(u-\b_+)\bar{A}(u),\label{a-function-6}\\
\bar{d}(u)&=&\bar{a}(-u-\eta).\label{d-function-6}
\eea

\subsection{$T-Q$ ansatz for even $N$}
Motivated by the results of  \cite{Cao13,Cao13-1}, we introduce
\footnote{Some deformed $T-Q$ ansatz for the eigenvalues of the
graded XXZ open chain was introduced  in \cite{Kar13}.} \bea
\L(u)&=& \bar{a}(u)\frac{Q_1(u-\eta)}{Q_2(u)}+
\bar{d}(u)\frac{Q_2(u+\eta)}{Q_1(u)}\no\\
&&\quad\quad +\frac{2\bar{c}\sinh(2u)\sinh(2u+2\eta)}{Q_1(u)Q_2(u)}\bar{A}(u)\bar{A}(-u-\eta), \label{T-Q-6-even}
\eea
where the functions $Q_1(u)$ and $Q_2(u)$ are
some trigonometric polynomials parameterized by $N$  Bethe roots
$\{\mu_j|j=1,\ldots,N\}$ as follows,
\bea
 Q_1(u)&=&\prod_{j=1}^{N}\frac{\sinh(u-\mu_j)}
 {\sinh(\eta)},\label{Q-1-6}\\
 Q_2(u)&=&\prod_{j=1}^{N}\frac{\sinh(u+\mu_j+\eta)}
 {\sinh(\eta)}=Q_1(-u-\eta).\label{Q-2-6}
\eea the parameters $\bar{c}$ is determined by the boundary
parameters and $\mu_j$ \bea
\bar{c}=\cosh((N+1)\eta+\a_-+\b_-+\a_++\b_+\hspace{-0.08truecm}+\hspace{-0.08truecm}
2\sum_{j=1}^N\mu_j)
\hspace{-0.08truecm}-\hspace{-0.08truecm}\cosh(\theta_-\hspace{-0.08truecm}-\hspace{-0.08truecm}\theta_+).
\label{c-para-1} \eea
The above relation leads to the fact that the
asymptotic behavior (\ref{Eigen-Asy-6}) of the eigenvalue
$\Lambda(u)$ is automatically satisfied. Let us evaluate the function (\ref{T-Q-6-even}) at points $\theta_j$ and
$\theta_j-\eta$
\bea
\Lambda(\theta_j)=\bar{a}(\theta_j)\frac{Q_1(\theta_j-\eta)}{Q_2(\theta_j)},\quad
\Lambda(\theta_j-\eta)=\bar{d}(\theta_j-\eta)\frac{Q_2(\theta_j)}{Q_1(\theta_j-\eta)},\quad j=1,\ldots,N, \no
\eea which yields that
\bea
\Lambda(\theta_j)\Lambda(\theta_j-\eta)=\bar{a}(\theta_j)\bar{d}(\theta_j-\eta),\quad j=1,\ldots,N.\no
\eea This implies that the function $\L(u)$ indeed satisfies the required identities (\ref{Eigen-Identity-6-1}).
If the $N$ parameters
$\{\mu_j|j=1,\ldots,N\}$ satisfy the following Bethe ansatz
equations (BAEs) \bea
&&\frac{2\bar{c}\sinh(2\mu_j)\sinh(2\mu_j+2\eta)\,\bar{A}(\mu_j)\bar{A}(-\mu_j-\eta)}
{\bar{d}(\mu_j)Q_2(\mu_j)Q_2(\mu_j+\eta)}=-1,
\,\,j=1,\ldots,N,\label{BAE-6} \eea with the following selection
rule for the roots of the above equations \bea \mu_j\neq \mu_l\quad
{\rm and}\quad \mu_j\neq -\mu_l-\eta, \label{selection} \eea the
function $\L(u)$ becomes the solution of
(\ref{Eigen-Identity-6-1})-(\ref{Eigen-Anal-6}).

The Bethe ansatz equation (\ref{BAE-6}) in the homogeneous limit
$\theta_j=0$ reads
\bea
&&\frac{\bar{c}\sinh(2\mu_j+\eta)\sinh(2\mu_j+2\eta)} {2
\sinh(\mu_j+\a_-+\eta)\cosh(\mu_j+\b_-+\eta)\sinh(\mu_j+\a_++\eta)\cosh(\mu_j+\b_++\eta) }\no\\
&&\qquad= \prod_{l=1}^N
\frac{\sinh(\mu_j+\mu_l+\eta)\sinh(\mu_j+\mu_l+2\eta)}{\sinh(\mu_j+\eta)\sinh(\mu_j+\eta)},
\,\,j=1,\ldots,N.\label{B1AE-6z} \eea The eigenvalue of the
Hamiltonian is given by \bea
&&E=-\sinh\eta[\coth(\alpha_-)+\tanh(\beta_-)+\coth(\alpha_+)+\tanh(\beta_+)]
\no
\\ &&\qquad\qquad-2\sinh\eta
\sum_{j=1}^{N}\coth(\mu_j+\eta) +(N-1)\cosh\eta.\label{2Spectrumtt}
\eea

Some remarks are in order. In \cite{Cao13-1}, a more general
form of $T-Q$ ansatz was proposed, which involves parameters $\l_j$,
$\mu_j$ and $\nu_j$. However, the numerical analysis of the
solutions of the associated Bethe ansatz equations (the rational
version of the Bethe ansatz equation (\ref{BAE-6})) for some small
sites $N$ strongly suggests that a fixed $M$ may give a complete set
of solutions of the transfer matrix. In such a sense, different $M$
in \cite{Cao13-1} just give different parametrization of the
eigenvalues but not different states. This suggests us in this paper
to adopt  the above parametrization of the $T-Q$ relations
(\ref{T-Q-6-even}) and  the following ones such as
(\ref{T-Q-6-odd}), (\ref{T-Q-8-even}) and (\ref{T-Q-8-odd}). Moreover,
numerical solutions of the BAEs (\ref{BAE-6}) for small size with random choices
of $\eta$ and the boundary parameters $\a_{\pm}$,
$\b_{\pm}$ and $\theta_{\pm}$ strongly suggest that the BAEs would give
the complete solutions of the model (namely, the eigenvalues calculated
from the BAEs coincide exactly to those obtained from exact
diagonalization). The numerical results for the $N=4$ case with the parameters:
$\eta=0.5$, $\alpha_+=1$, $\alpha_-=0.8$,
$\beta_+=0.4$, $\beta_-=0.3$, $\theta_+=0.7i$ and $\theta_-=0.9i$ are
shown in TABEL 1.
{\tiny
\begin{table}
\caption{\label{N4}
Numerical solutions of the BAEs for the $N=4$ case with the parameters:
$\eta=0.5$, $\alpha_+=1$, $\alpha_-=0.8$,
$\beta_+=0.4$, $\beta_-=0.3$, $\theta_+=0.7i$ and $\theta_-=0.9i$. $E$ is the eigevalues of the Hamiltonian.
The eigenvalues are exactly the same as those from the exact diagonalization.
}\vspace{0.6truecm}
\small
\begin{tabular}{ cccc|c|c} \hline \hline
$\mu_1$ & $\mu_2$ & $\mu_3$ & $\mu_4$ & $E$ & $n$ \\
\hline
$-0.3330-0.4622i$ & $-0.3330+0.4622i$ & $-0.2506-0.1242i$ & $-0.2506+0.1242i$ & $-6.8670$ & $1$ \\
$-1.0988-1.5708i$ & $-0.6931+1.5708i$ & $-0.2500-0.1095i$ & $-0.2500+0.1095i$ & $-4.8468$ & $2$ \\
$-1.0171+0.0000i$ & $-0.2501-0.0969i$ & $-0.2501+0.0969i$ & $-0.2457+1.5708i$ & $-3.9266$ & $3$ \\
$-2.0689-1.5708i$ & $-1.1954-0.0000i$ & $-0.2500-0.0945i$ & $-0.2500+0.0945i$ & $-3.2170$ & $4$ \\
$-1.1075-1.5708i$ & $-0.6883+1.5708i$ & $-0.2500-0.2762i$ & $-0.2500+0.2762i$ & $-1.6077$ & $5$ \\
$-1.0558-0.0000i$ & $-0.2497-0.2356i$ & $-0.2497+0.2356i$ & $-0.2458-1.5708i$ & $-1.2212$ & $6$ \\
$-2.1748+1.5708i$ & $-1.2064+0.0000i$ & $-0.2499-0.2179i$ & $-0.2499+0.2179i$ & $-0.6645$ & $7$ \\
$-0.6387-0.5472i$ & $-0.6387+0.5472i$ & $-0.1043+0.5711i$ & $-0.1043+2.5705i$ & $0.5747$ & $8$ \\
$-1.2297-1.5708i$ & $-0.5885+1.5708i$ & $-0.2591+0.9163i$ & $-0.2591+2.2253i$ & $1.5474$ & $9$ \\
$-1.1140+0.0000i$ & $-0.2950-0.6757i$ & $-0.2950+0.6757i$ & $0.1594-1.5708i$ & $1.8547$ & $10$ \\
$-3.2361-1.5708i$ & $-1.2319-0.0000i$ & $-0.2509-0.4386i$ & $-0.2509+0.4386i$ & $2.0475$ & $11$ \\
$-2.9114-3.1416i$ & $-0.9120-1.5708i$ & $-0.7921+1.5708i$ & $1.9542+1.5708i$ & $2.2990$ & $12$ \\
$-1.1848+0.0000i$ & $-1.0107+1.5708i$ & $-0.7379-1.5708i$ & $0.7058-0.0000i$ & $2.8030$ & $13$ \\
$-1.1547+0.0000i$ & $-0.9587-0.0000i$ & $-0.4986-1.5708i$ & $-0.0459+0.0000i$ & $3.3542$ & $14$ \\
$-1.5932+0.0000i$ & $-1.2867+0.0000i$ & $-0.4656+1.5708i$ & $0.1759-1.5708i$ & $3.8081$ & $15$ \\
$-1.5191-0.0000i$ & $-1.2970-0.0000i$ & $-1.0092-1.5708i$ & $0.8832+1.5708i$ & $4.0622$ & $16$ \\
\hline\hline \end{tabular}
\end{table}
}

It follows from (\ref{c-para-1}) that the parameter $c$ does depend up not
only the boundary parameters but also the parameters $\{\mu_j\}$ (such
a dependence on the parameters $\{\mu_j\}$  also
appears in the open  XYZ chain (see section 5 below)). The vanishing
condition of $c$, i.e. $c=0$, will leads to the constraint among the boundary parameters
found in \cite{Cao03,Yan07}, where one could find a proper ``local vacuum" to
apply the conventional Bethe ansatz. The Bethe ansatz equations
(\ref{B1AE-6z}) imply that for this case the parameters $\{\mu_j\}$ have to form
two types of pairs:
\bea (\mu_l,-\mu_l-\eta),\quad\quad
(\mu_l,-\mu_l-2\eta).\no
\eea Suppose  the number of the first type
pairs is $M$, the resulting $T-Q$ relation (\ref{T-Q-6-even}) becomes the
conventional one \bea
\L(u)=\bar{a}(u)\frac{\bar{Q}(u-\eta)}{\bar{Q}(u)}+
\bar{d}(u)\frac{\bar{Q}(u+\eta)}{\bar{Q}(u)},\label{New-1} \eea with
\bea
\bar{Q}(u)=\prod_{j=1}^M\frac{\sinh(u-\mu_l)\sinh(u+\mu_l+\eta)}{\sinh^2\eta}.\no
\eea The constraint (\ref{c-para-1}) allows us to fix the integer $M$
by the following condition which the boundary parameters must obey
\bea
(N-1-2M)\eta=\a_-+\b_-+\a_++\b_+\pm(\theta_--\theta_+)=k\eta,\quad \mbox{mod}(i\pi). \label{Constraint-1}
\eea It
is exactly the constrained boundary parameters for which a proper
``local vacuum" exists \cite{Cao03}. If the boundary parameter obey the constraint
(\ref{Constraint-1}), there exists another solution to the Bethe ansatz equations
(\ref{B1AE-6z}) which also corresponds to $c=0$
\bea
 (\mu_l,-\mu_l-\eta),\quad (-\a_--\eta, -\a_+-\eta, -\b_--\eta+\frac{i\pi}{2},
 -\b_+-\eta+\frac{i\pi}{2}), \quad (\mu_l,-\mu_l-2\eta).\no
\eea Now let the number of the first type pairs be $\bar{M}$, the constraint
(\ref{c-para-1}) allows us to fix the integer $\bar{M}$ by
\bea
(-N+1+2\bar{M})\eta=\a_-+\b_-+\a_++\b_+\pm(\theta_--\theta_+)=k\eta,\quad \mbox{mod}(i\pi). \label{Constraint-2}
\eea The resulting $T-Q$ relation (\ref{T-Q-6-even}) becomes another
conventional one (cf. (\ref{New-2}))
\bea
\L(u)=\tilde{\bar{a}}(u)\frac{\bar{Q}(u-\eta)}{\bar{Q}(u)}+
\tilde{\bar{d}}(u)\frac{\bar{Q}(u+\eta)}{\bar{Q}(u)},\label{New-2} \eea
with
\bea
&&\tilde{\bar{a}}(u)=-2^2\frac{\sinh(2u+2\eta)}{\sinh(2u+\eta)}\sinh(u+\a_-)\cosh(u+\b_-)\no\\
&&\quad\quad\quad\quad\times\sinh(u+\a_+)\cosh(u+\b_+)\bar{A}(u),
\quad \tilde{\bar{d}}(u)=\tilde{\bar{a}}(-u-\eta),\no\\
&&\bar{Q}(u)=\prod_{j=1}^{\bar{M}}\frac{\sinh(u-\mu_l)\sinh(u+\mu_l+\eta)}{\sinh^2\eta}.\no
\eea Then the two resulting conventional T-Q relations (\ref{New-1}) and (\ref{New-2}) with the numbers constraints
(\ref{Constraint-1}) and (\ref{Constraint-2}) recover the Bethe ansatz solutions
\cite{Nep03,Yan06} of the open XXZ chain when boundary parameters satisfy the constraint (\ref{Constraint-1}).
Moreover, one can check that
when the K-matrices are diagonal ones which correspond to the cases
of $\a_{\pm}$ or $\b_{\pm}$ $\longrightarrow \infty$, there is no
constrain for the choices of the integer $M$. Hence the resulting
$T-Q$ ansatz  is reduced to the usual form parameterized by
a discrete $M=0,\ldots,N$.


\subsection{$T-Q$ ansatz for odd $N$}
For the case of $N$ being odd, we introduce \bea \L(u)&=&
\bar{a}(u)\frac{Q_1(u-\eta)}{Q_2(u)}+
\bar{d}(u)\frac{Q_2(u+\eta)}{Q_1(u)}\no\\
&+&\frac{2\bar{c}\sinh(2u)\sinh(2u+2\eta)}{Q_1(u)Q_2(u)}
\frac{\sinh u\sinh(u+\eta)}{\sinh^2\eta}\bar{A}(u)\bar{A}(-u-\eta),
\label{T-Q-6-odd} \eea where the functions $Q_1(u)$ and $Q_2(u)$ are
some trigonometric polynomials parameterized by $N+1$  Bethe roots
$\{\mu_j|j=1,\ldots,N+1\}$ as follows, \bea
 Q_1(u)&=&\prod_{j=1}^{N+1}\frac{\sinh(u-\mu_j)}
 {\sinh(\eta)},\label{Q-3-6}\\
 Q_2(u)&=&\prod_{j=1}^{N+1}\frac{\sinh(u+\mu_j+\eta)}
 {\sinh(\eta)}=Q_1(-u-\eta).\label{Q-4-6}
\eea
the parameters $\bar{c}$ is determined by the boundary parameters and $\mu_j$
\bea
\bar{c}=\cosh((N+3)\eta+\a_-+\b_-+\a_++\b_+\hspace{-0.08truecm}+\hspace{-0.08truecm}
2\sum_{j=1}^{N+1}\mu_j)
\hspace{-0.08truecm}-\hspace{-0.08truecm}\cosh(\theta_-\hspace{-0.08truecm}-\hspace{-0.08truecm}\theta_+).
\label{c-para-2}
\eea The above relation leads to the fact that the asymptotic behavior (\ref{Eigen-Asy-6}) of the eigenvalue $\Lambda(u)$
is automatically satisfied.  Then the function $\L(u)$ given by (\ref{T-Q-6-odd}) becomes the solution of
(\ref{Eigen-Identity-6-1})-(\ref{Eigen-Anal-6}) provided that
the $N+1$ parameters $\{\mu_j|j=1,\ldots,N+1\}$
satisfy the following Bethe ansatz equations
\bea
&&\frac{2\bar{c}\sinh(2\mu_j)\sinh(2\mu_j+2\eta)\,\bar{A}(\mu_j)\bar{A}(-\mu_j-\eta)}
{\bar{d}(\mu_j)Q_2(\mu_j)Q_2(\mu_j+\eta)}=
-\frac{\sinh^2\eta}{\sinh\mu_j\sinh(\mu_j+\eta)},\no\\
&&\quad\quad \,j=1,\ldots,N+1,\label{BAE-7} \eea with the very
selection rule (\ref{selection}) for the roots of the above
equations. In the homogeneous limit $\theta_j=0$, the Bethe ansatz
equations (\ref{BAE-7}) can be written as
\bea
&&\frac{\bar{c}\sinh(2\mu_j+\eta)\sinh(2\mu_j+2\eta)\sinh\mu_j\sinh(\mu_j+\eta)\,\sinh^{2N}(\mu_j+\eta)}
{2\sinh(\mu_j+\a_-+\eta)\cosh(\mu_j+\b_-+\eta)\sinh(\mu_j+\a_++\eta)\cosh(\mu_j+\b_++\eta) }\no\\
&&\qquad= \prod_{l=1}^{N+1}
\sinh(u_j+u_l+\eta)\sinh(u_j+u_l+2\eta),
\,\,j=1,\ldots, N+1.\label{B1AE-7z} \eea The eigenvalue of the
Hamiltonian reads \bea
&&E=-\sinh\eta[\coth(\alpha_-)+\tanh(\beta_-)+\coth(\alpha_+)+\tanh(\beta_+)]
\no
\\ &&\qquad\qquad-2\sinh\eta
\sum_{j=1}^{N+1}\coth(\mu_j+\eta)
+(N-1)\cosh\eta.\label{2Spectrumtt2} \eea

Numerical solutions of the BAEs (\ref{B1AE-7z}) for small size with random choices
of $\eta$ and the boundary parameters $\a_{\pm}$,
$\b_{\pm}$ and $\theta_{\pm}$ strongly suggest that the BAEs might give
the complete solutions of the model.
The numerical results for the $N=3$ case with the parameters:
$\eta=0.5$, $\alpha_+=1$, $\alpha_-=0.8$,
$\beta_+=0.4$, $\beta_-=0.3$, $\theta_+=0.7i$ and $\theta_-=0.9i$ are
shown in TABEL 2.
{\tiny
\begin{table}
\caption{\label{N3}
Numerical solutions of the BAEs for the $N=3$ case with the parameters:
$\eta=0.5$, $\alpha_+=1$, $\alpha_-=0.8$,
$\beta_+=0.4$, $\beta_-=0.3$, $\theta_+=0.7i$ and $\theta_-=0.9i$. $E$ is the eigevalues of the Hamiltonian.
The eigenvalues are exactly the same as those from the exact diagonalization.
}
\small
\begin{center}
\begin{tabular}{ cccc|c|c} \hline \hline
$\mu_1$ & $\mu_2$ & $\mu_3$ & $\mu_4$ & $E$ & $n$ \\
\hline
$-0.5276-0.3652i$ & $-0.5276+0.3652i$ & $-0.2481-0.1756i$ & $-0.2481+0.1756i$ & $-4.8590$ & $1$ \\
$-2.9056+0.0000i$ & $-1.1969-0.0000i$ & $-0.2500-0.1261i$ & $-0.2500+0.1261i$ & $-3.5939$ & $2$ \\
$-0.6974-0.5166i$ & $-0.6974+0.5166i$ & $-0.2826-0.4381i$ & $-0.2826+0.4381i$ & $-0.1251$ & $3$ \\
$-0.9296-0.0000i$ & $-0.2637-0.4333i$ & $-0.2637+0.4333i$ & $0.6919+1.5708i$ & $-0.0479$ & $4$ \\
$-0.9741+4.7124i$ & $-0.7424-4.7124i$ & $-0.5001-0.5664i$ & $-0.5001+0.5664i$ & $1.1449$ & $5$ \\
$-1.1498+0.0000i$ & $-0.5212-2.5079i$ & $-0.5212-0.6337i$ & $0.0230+1.5708i$ & $1.8855$ & $6$ \\
$-1.1060-0.1659i$ & $-1.1060+0.1659i$ & $-0.5216-1.5708i$ & $0.3535+0.0000i$ & $2.5676$ & $7$ \\
$-1.5205+0.0000i$ & $-1.2965-0.0000i$ & $-1.1030+1.5708i$ & $0.8003-1.5708i$ & $3.0278$ & $8$ \\
\hline\hline \end{tabular} \end{center}
\end{table}
}


\section{Results for the open XYZ chain}
\label{Hom} \setcounter{equation}{0}
\subsection{Operator identity}
Now  we consider the most general solutions $K^{\mp}(u)$
\cite{Ina94} of the reflection equation and its dual equation
associated with the eight-vertex R-matrix given by
(\ref{Eight-1})-(\ref{Eight-2}),

\bea
K^{-}(u)&=&\hspace{-0.2cm}\frac{\s(2u)}{2\s(u)}\lt\{{\rm id}\hspace{-0.1cm}
+\hspace{-0.1cm} \frac{c^{(-)}_x\s(u)e^{-i\pi
u}}{\s(u+\frac{\tau}{2})}\s^x\hspace{-0.1truecm}
+\hspace{-0.1cm}\frac{c^{(-)}_y\s(u)e^{-i\pi
u}}{\s(u+\frac{1+\tau}{2})}\s^y  \hspace{-0.1cm}+ \hspace{-0.1cm}
\frac{c^{(-)}_z\s(u)}{\s(u+\frac{1}{2})}\s^z
\rt\},\label{K-matrix1}\\
K^{+}(u)&=&\hspace{-0.2cm}\lt.K^-(-u-\eta)\rt|_{\{c^{(-)}_l\}\rightarrow
\{c^{(+)}_l\}},\label{K-matrix2} \eea where the constants $\{c^{(\mp)}_l\}$ are expressed
in terms of boundary parameters $\{\a^{(\mp)}_l\}$ as follows:
\bea c^{(\mp)}_x&=&e^{-i\pi(\sum_l\a^{(\mp)}_l-\frac{\tau}{2})}
\prod_{l=1}^3\frac{\s(\a^{(\mp)}_l-\frac{\tau}{2})}
{\s(\a^{(\mp)}_l)}, \quad  c^{(\mp)}_z=
\prod_{l=1}^3\frac{\s(\a^{(\mp)}_l-\frac{1}{2})}
{\s(\a^{(\mp)}_l)},\no\\
c^{(\mp)}_y&=&e^{-i\pi(\sum_l\a^{(\mp)}_l-\frac{1}{2}-\frac{\tau}{2})}
\prod_{l=1}^3\frac{\s(\a^{(\mp)}_l-\frac{1}{2}-\frac{\tau}{2})}
{\s(\a^{(\mp)}_l)}.\eea
Direct calculation \cite{Yan08-e} shows that
\bea
{\rm Det}_q(K^-(u))&=&\frac{\s(2u-2\eta)}{\s(\eta)}\prod_{l=1}^3\frac{\s(\a^{(-)}_l+u)\s(\a^{(-)}_l-u)}
{\s(\a^{(-)}_l)\s(\a^{(-)}_l)},\\
{\rm Det}_q(K^+(u))&=&-\frac{\s(2u+2\eta)}{\s(\eta)}\prod_{l=1}^3\frac{\s(\a^{(+)}_l+u)\s(\a^{(+)}_l-u)}
{\s(\a^{(+)}_l)\s(\a^{(+)}_l)}.
\eea This leads to that for the eight-vertex model the function $\delta(u)$ defined by (\ref{Det-function-1}) reads
\bea
\delta(u)&=&-\frac{\s(2u+2\eta)\s(2u-2\eta)}{\s(\eta)\,\s(\eta)}
\prod_{\g=\pm}\prod_{l=1}^3\frac{\s(u+\a^{(\g)}_l)\s(u-\a^{(\g)}_l)}
{\s(\a^{(\g)}_l)\s(\a^{(\g)}_l)}\no\\
&&\times\prod_{l=1}^N\frac{\s(u+\theta_l+\eta)\s(u+\theta_l-\eta)\s(u-\theta_l+\eta)\s(u-\theta_l-\eta)}
{\s(\eta)\s(\eta)\s(\eta)\s(\eta)}.\label{delta-function-elliptic}
\eea Following the method in \cite{Yan08-e,Beh96} and using the
crossing relation of the R-matrix (\ref{crosing-unitarity}) and the
explicit expressions of the K-matrices (\ref{K-matrix1}) and
(\ref{K-matrix2}), one can show that the corresponding transfer
matrix $t(u)$ satisfies the following crossing relation \bea
  t(-u-\eta)=t(u).\label{transfer-crosing-1}
\eea The quasi-periodicity of $\s$-function
(\ref{quasi-func}) allows one to derive the
following properties of the R-matrix and K-matrices:
\bea
R_{12}(u+1)&=&-\s^z_1R_{12}(u)\s^z_1=-\s^z_2R_{12}(u)\s^z_2,\quad
K^{\mp}(u+1)=-\s^zK^{\mp}(u)\s^z,\label{quasi-periodic-R-1}\\
R_{12}(u+\tau)&=&-e^{-2i\pi(u+\frac{\eta}{2}+\frac{\tau}{2})}
\s^x_1R_{12}(u)\s^x_1=-e^{-2i\pi(u+\frac{\eta}{2}+\frac{\tau}{2})}
\s^x_2R_{12}(u)\s^x_2,\label{quasi-periodic-R-2}\\
R_{12}(u+1+\tau)&=&e^{-2i\pi(u+\frac{\eta}{2}+\frac{\tau}{2})}
\s^y_1R_{12}(u)\s^y_1=e^{-2i\pi(u+\frac{\eta}{2}+\frac{\tau}{2})}
\s^y_2R_{12}(u)\s^y_2,\label{quasi-periodic-R-3}\\
K^-(u+\tau)&=&-e^{-2i\pi(3u+\frac{3}{2}\tau)}\s^xK^-(u)\s^x,\label{quasi-periodic-K-1}\\
K^-(u+1+\tau)&=&e^{-2i\pi(3u+\frac{3}{2}\tau)}\s^yK^-(u)\s^y,\label{quasi-periodic-K-2}\\
K^+(u+\tau)&=&-e^{-2i\pi(3u+3\eta+\frac{3}{2}\tau)}\s^xK^+(u)\s^x,\label{quasi-periodic-K-3}\\
K^+(u+\tau)&=&e^{-2i\pi(3u+3\eta+\frac{3}{2}\tau)}\s^yK^+(u)\s^y.
\label{quasi-periodic-K-4} \eea From these relations one obtains the
quasi-periodic properties of the transfer matrix $t(u)$, \bea
t(u+1)=t(u),\quad
t(u+\tau)=e^{-2i\pi(N+3)(2u+\eta+\tau)}\,t(u).\label{trans-Perio}
\eea With the help of the explicit expression (\ref{K-matrix1}) of
the K-matrix $K^-(u)$, one may derive that \bea
K^-(0)&=&\frac{1}{2}tr(K^-(0))\times {\rm id},\quad K^-(\frac{1}{2})=\frac{1}{2}tr(K^-(\frac{1}{2})\s^z)\times \s^z,\label{K-value-1}\\
K^-(\frac{\tau}{2})&=& \frac{1}{2}tr(K^-(\frac{\tau}{2})\s^x)\times \s^x,\quad
K^-(\frac{1+\tau}{2})= \frac{1}{2}tr(K^-(\frac{1+\tau}{2})\s^y)\times \s^y.\label{K-value-2}
\eea Then we can evaluate the transfer matrix $t(u)$ at these particular points
\bea
t(0)&=&\frac{1}{2}tr(K^+(0))tr(K^-(0))\prod_{l=1}^N\frac{\s(\eta+\theta_l)\s(\eta-\theta_l)}{\s(\eta)\s(\eta)}\times {\rm id},\label{t-0}\\
t(\frac{1}{2})&=&\frac{1}{2}tr(K^+(\frac{1}{2})\s^z)tr(K^-(\frac{1}{2})\s^z)(-1)^N\no\\
&&\times\prod_{l=1}^N\frac{\s(\eta+\frac{1}{2}+\theta_l)
\s(\eta-\frac{1}{2}-\theta_l)}{\s(\eta)\s(\eta)}\times {\rm id},\label{t-1}\\
t(\frac{\tau}{2})&=&\frac{1}{2}tr(K^+(\frac{\tau}{2})\s^x)tr(K^-(\frac{\tau}{2})\s^x)(-1)^Ne^{-2\pi i\{\frac{N}{2}\eta-\sum_{j=1}^N\theta_j\}}\no\\
&&\times\prod_{l=1}^N\frac{\s(\eta+\frac{\tau}{2}+\theta_l)
\s(\eta-\frac{\tau}{2}-\theta_l)}{\s(\eta)\s(\eta)}\times {\rm id},\label{t-2}\\
t(\frac{1+\tau}{2})&=&\frac{1}{2}tr(K^+(\frac{1+\tau}{2})\s^y)tr(K^-(\frac{1+\tau}{2})\s^y)e^{-2\pi i\{\frac{N}{2}\eta-\sum_{j=1}^N\theta_j\}}(-1)^N\no\\
&&\times\prod_{l=1}^N\frac{\s(\eta+\frac{1+\tau}{2}+\theta_l)
\s(\eta-\frac{1+\tau}{2}-\theta_l)}{\s(\eta)\s(\eta)}\times {\rm id}.\label{t-3}
\eea

\subsection{Functional relations}

The very operator identity (\ref{Operator-id-3}) implies the corresponding eigenvalue
$\Lambda(u)$ also satisfies the similar relations
\bea
\Lambda(\theta_j)\Lambda(\theta_j-\eta)=\frac{\delta(\theta_j)\,\s(\eta)\,\s(\eta)}{\s(\eta-2\theta_j)\,\s(\eta+2\theta_j)},
\quad j=1,\ldots,N, \label{Eigen-Identity-8-1}
\eea where for the XYZ open spin chain the function $\delta(u)$ is given by (\ref{delta-function-elliptic}).
The properties
of the transfer matrix $t(u)$ given by
(\ref{transfer-crosing-1}) and (\ref{t-0})-(\ref{t-3}) imply that the corresponding eigenvalue
$\Lambda(u)$ satisfies the following relations:
\bea
\L(-u-1)&=&\L(u),\label{Eig-Cro}\\
\Lambda(0)&=&\frac{1}{2}tr(K^+(0))tr(K^-(0))\prod_{l=1}^N\frac{\s(\eta+\theta_l)\s(\eta-\theta_l)}{\s(\eta)\s(\eta)},\label{Eigen-8-0}\\
\L(\frac{1}{2})&=&\frac{1}{2}tr(K^+(\frac{1}{2})\s^z)tr(K^-(\frac{1}{2})\s^z)(-1)^N\no\\
&&\times\prod_{l=1}^N\frac{\s(\eta+\frac{1}{2}+\theta_l)
\s(\eta-\frac{1}{2}-\theta_l)}{\s(\eta)\s(\eta)},\label{Eigen-8-1}\\
\L(\frac{\tau}{2})&=&\frac{1}{2}tr(K^+(\frac{\tau}{2})\s^x)tr(K^-(\frac{\tau}{2})\s^x)(-1)^Ne^{-2\pi i\{\frac{N}{2}\eta-\sum_{j=1}^N\theta_j\}}\no\\
&&\times\prod_{l=1}^N\frac{\s(\eta+\frac{\tau}{2}+\theta_l)
\s(\eta-\frac{\tau}{2}-\theta_l)}{\s(\eta)\s(\eta)},\label{Eigen-8-2}\\
\L(\frac{1+\tau}{2})&=&\frac{1}{2}tr(K^+(\frac{1+\tau}{2})\s^y)tr(K^-(\frac{1+\tau}{2})\s^y)e^{-2\pi i\{\frac{N}{2}\eta-\sum_{j=1}^N\theta_j\}}(-1)^N\no\\
&&\times\prod_{l=1}^N\frac{\s(\eta+\frac{1+\tau}{2}+\theta_l)
\s(\eta-\frac{1+\tau}{2}-\theta_l)}{\s(\eta)\s(\eta)}.\label{Eigen-8-3}
\eea The quasi-periodic properties (\ref{trans-Perio}) of the
transfer matrix $t(u)$ allow us to derive the following
quasi-periodic properties of the corresponding eigenvalue \bea
\L(u+1)=\L(u),\quad
\L(u+\tau)=e^{-2i\pi(N+3)(2u+\eta+\tau)}\,\L(u).\label{Eigen-Perio}
\eea The analyticity of the R-matrix and K-matrices and independence
on $u$ of the eigenstate lead to that the eigenvalue $\Lambda(u)$
further possesses  the property \bea \L(u) \mbox{, as an entire
function of $u$, is an elliptic polynomial of degree
$2N+6$}.\label{Eigen-Anal-8} \eea Therefore the values of $\L(u)$ at
generic $2N+6$ points in the fundamental domain of the elliptic
functions suffice to determine  the function uniquely. Actually, we
have already obtained the corresponding values of  $\L(u)$ at points
$u=0,\,\frac{1}{2},\,\frac{\tau}{2}$, as well as their crossing
points $-\eta,\, -\frac{1}{2}-\eta,\,-\frac{\tau}{2}-\eta$ via the
relation (\ref{Eig-Cro}). With the help of the relation
(\ref{Eigen-Identity-8-1}), one can further obtain the values of
$\L(u)$ at other $2N$ points $\{\theta_j|j=1,\ldots,N\}$ and their
crossing points $\{-\theta_j-\eta|j=1,\ldots,N\}$.  Then we can
completely determine the eigenvalue function $\Lambda(u)$.

Let us introduce some functions $A(u)$, $a(u)$ and $d(u)$
\bea
A(u)&=& \prod_{j=1}^{N}\frac{\s(u+\theta_j+\eta)\,\s(u-\theta_j+\eta)}{\s(\eta)\,\s(\eta)}, \\
a(u)&=&-\frac{\s(2u+2\eta)}{\s(2u+\eta)}
\prod_{\g=\pm} \prod_{l=1}^3
\frac{\s(u-\a_l^{(\g)})}
{\s(\a_l^{(\g)})}A(u),\label{a-function-8}\\
d(u)&=&a(-u-\eta).\label{d-function-8}
\eea

\subsubsection{$T-Q$ ansatz for even $N$}
For the case of $N$ being even, we can construct the solutions of (\ref{Eigen-Identity-8-1})-(\ref{Eigen-Anal-8}) by the following ansatz
\bea
\Lambda(u)&=& a(u)\frac{Q_1(u-\eta)}{Q_2(u)}+
 d(u)\frac{Q_2(u+\eta)}{Q_1(u)}\no\\
&&+\frac{c\,\s(2u)\s(2u+2\eta)}{Q_1(u)Q_2(u)}A(u)A(-u-\eta).
\label{T-Q-8-even}
\eea The functions $Q_1(u)$ and $Q_2(u)$ are parameterized by $N+1$  Bethe roots
$\{\mu_j|j=1,\ldots,N+1\}$ as follows,
\bea
 Q_1(u)&=&\prod_{j=1}^{N+1}\frac{\s(u-\mu_j)}
 {\s(\eta)},\label{Q-1-8}\\
 Q_2(u)&=&\prod_{j=1}^{N+1}\frac{\s(u+\mu_j+\eta)}
 {\s(\eta)}.\label{Q-2-8}
\eea These $N+1$ parameters $\mu_j$ (which are different from each other) and $c$   should satisfy
the following $N+2$ equations
\bea
&&\sum_{\g=\pm}\sum_{l=1}^3\a^{(\g)}_l+(N+3)\eta+2\sum_{j=1}^{N+1}\mu_j=0\,\,{\rm mod} (1),\label{BAE-8-1}\\[6pt]
&&\frac{c\,\s(2\mu_j)\s(2\mu_j+2\eta)\,A(\mu_j)A(-\mu_j-\eta)}
{d(\mu_j)Q_2(\mu_j)Q_2(\mu_j+\eta)}=-1,\,j=1,\ldots, N+1,\quad
\label{BAE-8-2} \eea with the very selection rule (\ref{selection})
for the roots of the above equations. In the homogeneous limit, the
Bethe ansatz equations (\ref{BAE-8-2}) become
\bea
&&c\,\s^2(\eta)\s(2\mu_j+\eta)\s(2\mu_j+2\eta)\s^{2N}(\mu_j+\eta)
\no \\
&&\qquad = \prod_{\g=\pm} \prod_{l=1}^3
\frac{\s(\mu_j+\a_l^{(\g)}+\eta)} {\s(\a_l^{(\g)})}
\prod_{l=1}^{N+1}\s(\mu_j+\mu_l+\eta)\s(\mu_j+\mu_l+2\eta),\no \\
&&\qquad\qquad\qquad \qquad\qquad\qquad j=1,\ldots, N+1. \no \eea
The eigenvalues $E$ of the Hamiltonian (\ref{Ham}) read\bea
E\hspace{-0.1truecm}=\hspace{-0.1truecm}
\frac{\s(\eta)}{\s'(0)}\lt\{2\sum_{j=1}^{N+1}\lt(
\zeta(u_j)-\zeta(u_j+
\eta)\rt)\hspace{-0.1truecm}+\hspace{-0.1truecm}(N-1)\zeta(\eta)
+\sum_{\g=\pm}\sum_{l=1}^3\zeta(\a^{(\g)}_l)
\rt\}.\label{H-Eign}\eea

\subsubsection{$T-Q$ ansatz for odd $N$}
For the case of $N$ being odd, we need to construct the solutions of (\ref{Eigen-Identity-8-1})-(\ref{Eigen-Anal-8}) by the following ansatz
\bea
\Lambda(u)&=& a(u)\frac{Q_1(u-\eta)}{Q_2(u)}+
 d(u)\frac{Q_2(u+\eta)}{Q_1(u)}\no\\
&&+\frac{c\,\s(2u)\s(2u+2\eta)\s(u)\s(u+\eta)}{Q_1(u)Q_2(u)\s(\eta)\s(\eta)}A(u)A(-u-\eta).
\label{T-Q-8-odd}
\eea The functions $Q_1(u)$ and $Q_2(u)$ are parameterized by $N+2$  Bethe roots
$\{\mu_j|j=1,\ldots,N+2\}$ as follows,
\bea
 Q_1(u)&=&\prod_{j=1}^{N+2}\frac{\s(u-\mu_j)}
 {\s(\eta)},\label{Q-3-8}\\
 Q_2(u)&=&\prod_{j=1}^{N+2}\frac{\s(u+\mu_j+\eta)}
 {\s(\eta)}.\label{Q-4-8}
\eea These $N+2$ parameters $\mu_j$
(which are different from each other) and $c$   should satisfy
the following $N+3$ equations
\bea
&&\sum_{\g=\pm}\sum_{l=1}^3\a^{(\g)}_l+(N+5)\eta+2\sum_{j=1}^{N+1}\mu_j=0\,\,{\rm mod} (1),\label{BAE-8-3}\\[6pt]
&&\frac{c\,\s(2\mu_j)\s(2\mu_j+2\eta)\s(\mu_j)\s(\mu_j+\eta)A(\mu_j)A(-\mu_j-\eta)}
{d(\mu_j)Q_2(\mu_j)Q_2(\mu_j+\eta)\s(\eta)\s(\eta)}=-1,\,j=1,\ldots,
N+2,\quad \label{BAE-8-4} \eea with the very selection rule
(\ref{selection}) for the roots of the above equations. In the
homogeneous limit, the Bethe ansatz equations (\ref{BAE-8-4}) read
\bea
&&c\,\s^2(\eta)\s(\mu_j)\s(2\mu_j+\eta)\s(2\mu_j+2\eta)\s^{2N+1}(\mu_j+\eta)
\no \\
&&\qquad = \prod_{\g=\pm} \prod_{l=1}^3
\frac{\s(\mu_j+\a_l^{(\g)}+\eta)} {\s(\a_l^{(\g)})}
\prod_{l=1}^{N+2}\s(\mu_j+\mu_l+\eta)\s(\mu_j+\mu_l+2\eta),\no \\
&&\qquad\qquad\qquad \qquad\qquad\qquad j=1,\ldots, N+2. \no \eea
and the eigenvalue of the Hamiltonian is\bea
E\hspace{-0.1truecm}=\hspace{-0.1truecm}
\frac{\s(\eta)}{\s'(0)}\lt\{2\sum_{j=1}^{N+2}\lt(
\zeta(u_j)-\zeta(u_j+
\eta)\rt)\hspace{-0.1truecm}+\hspace{-0.1truecm}(N-1)\zeta(\eta)
+\sum_{\g=\pm}\sum_{l=1}^3\zeta(\a^{(\g)}_l)
\rt\}.\label{H-Eign1}\eea


\section{Conclusions}
\label{Con} \setcounter{equation}{0} The anisotropic spin-$\frac12$
chains with arbitrary boundary fields (i.e., there is not any
constrain to the boundary parameters, cf. \cite{Fan96,Cao03})
defined by (\ref{ohami}), which includes the most general open XXZ
chain and open XYZ chain, are studied by the off-diagonal Bethe
ansatz proposed in \cite{Cao13}. The eigenvalues of the transfer
matrix are given in terms of  generalized $T-Q$ ansatzs
(\ref{T-Q-6-even}), (\ref{T-Q-6-odd}), (\ref{T-Q-8-even}) and (\ref{T-Q-8-odd}). The
corresponding Bethe ansatz equations are given by (\ref{B1AE-6z}),
(\ref{B1AE-7z}), (\ref{BAE-8-1})-(\ref{BAE-8-2}),
(\ref{BAE-8-3})-(\ref{BAE-8-4}) respectively.

The different forms of BAEs  indicate different topological natures
for even $N$ and odd $N$ cases, which was firstly observed in the
closed XYZ chain. This phenomenon is also quite similar to that
appeared in the XXZ model, where the periodic and antiperiodic
boundary conditions also induce quite different Bethe ansatz
equations and indeed show different topological behaviors
\cite{Cao13}.

As for integrable models without $U(1)$ symmetry, most of
conventional Bethe ansatz methods failed because of the lack of a
proper ``local vacuum". The off-diagonal Bethe ansatz method
overcomes this obstacle by using functional relations to construct
the $T-Q$ ansatz, which is completely independent of the
representation basis and thus does not need any information of the
states. Although the functional relations between eigenvalues
$\Lambda(u)$ and the quantum determinant $\d(u)$ at some particular
points can be obtained in different ways for some special cases
\cite{Cao13,Cao13-1,Nic12,Fra11},
our $T-Q$ ansatz would play an important role to construct
manageable Bethe ansatz equations. In the present case, based on the
relation (\ref{Eigen-Identity-6-1}) (or (\ref{Eigen-Identity-8-1}))
and some properties (\ref{crosing-Eign-6})-(\ref{Eigen-Anal-6}) (or
(\ref{Eig-Cro})-(\ref{Eigen-Anal-8}) ) of $\Lambda(u)$ we can give a
generalized $T-Q$ ansatz, which modifies the usual $T-Q$ relation by
adding an extra off-diagonal term. Such an extra term encodes the
contribution of the off-diagonal element of the associated K-matrix.

In fact, the functional relation between the eigenvalue $\L(u)$ and
the quantum determinant $\d(u)$ is a direct consequence of the
operator identity (\ref{Operator-id-3}) which is obtained only via
some properties of the R-matrix and K-matrices. As we demonstrated,
similar operator relation holds for arbitrary integrable boundaries
(no matter periodic, anti-periodic or open boundaries). In such a
sense, our method would greatly simplify the process of Bethe ansatz
and would provide an unified procedure for approaching the
integrable models both with and without $U(1)$ symmetry.

\section*{Acknowledgments}

The financial support from  the National Natural Science Foundation
of China (Grant Nos. 11174335, 11075126, 11031005, 11375141,
11374334), the National Program for Basic Research of MOST (973
project under grant No.2011CB921700) and the State Education
Ministry of China (Grant No. 20116101110017 and SRF for ROCS) are
gratefully acknowledged. Two of the authors (W.-L. Yang and K. Shi)
would like to thank IoP/CAS for the hospitality and they enjoyed
during their visit there. We also would like to acknowledge
Y.Z. Jiang and S. Cui for their numerical helps.


\section*{Appendix A: Elliptic functions }
\setcounter{equation}{0}
\renewcommand{\theequation}{A.\arabic{equation}}

In this appendix, we give the definitions of elliptic functions
which appear in our study related to the XYZ models and some
identity relations between the functions.

Let us fix $\tau$ as such that ${\rm Im}(\tau)>0$. We introduce the following elliptic functions
\bea &&\theta\lt[
\begin{array}{c}
a\\b
\end{array}\rt](u,\tau)=\sum_{m=-\infty}^{\infty}
\exp\lt\{i\pi\lt[(m+a)^2\tau+2(m+a)(u+b)\rt]\rt\},\label{Function-a-b}\\
&&\s(u)=\t\lt[\begin{array}{c}\frac{1}{2}\\[2pt]\frac{1}{2}
\end{array}\rt](u,\tau),\quad \zeta(u)=\frac{\partial}{\partial u}
\lt\{\ln\s(u)\rt\}.\label{Function} \eea Among them the
$\s$-function\footnote{Our $\s$-function is the $\vartheta$-function
$\vartheta_1(u)$ in \cite{Whi50}. It has the following relation with
the {\it Weierstrassian\/} $\s$-function if one denotes it by
$\s_w(u)$: $\s_w(u)\propto e^{\eta_1u^2}\s(u)$,
$\eta_1=\pi^2(\frac{1}{6}-4\sum_{n=1}^{\infty}\frac{nq^{2n}}{1-q^{2n}})
$ and $q=e^{i\tau}$.} satisfies the Riemann-identity: \bea
&&\s(u+x)\s(u-x)\s(v+y)\s(v-y)-\s(u+y)\s(u-y)\s(v+x)\s(v-x)\no\\
&&~~~~~~=\s(u+v)\s(u-v)\s(x+y)\s(x-y),\label{identity}\eea and other
identities which have been used in this paper \bea
\s(2u)&=&\frac{2\s(u)\s(u+\frac{1}{2})\s(u+\frac{\tau}{2})
\s(u-\frac{1}{2}-\frac{\tau}{2})}{\s(\frac{1}{2})\s(\frac{\tau}{2})
\s(-\frac{1}{2}-\frac{\tau}{2})},\label{func-2}\\
\s(u+1)&=&-\s(u),\quad
\s(u+\tau)=-e^{-2i\pi(u+\frac{\tau}{2})}\s(u),\label{quasi-func}\\
\frac{\s(u)}{\s(\frac{\tau}{2})}&=&\frac{ \theta\lt[\begin{array}{l}
0\\\frac{1}{2}\end{array}\rt](u,2\tau)\,\,\theta\lt[\begin{array}{l}
\frac{1}{2}\\[2pt]\frac{1}{2}\end{array}\rt](u,2\tau)}
{\theta\lt[\begin{array}{l}
0\\\frac{1}{2}\end{array}\rt](\frac{\tau}{2},2\tau)\,\,
\theta\lt[\begin{array}{l}
\frac{1}{2}\\[2pt]\frac{1}{2}\end{array}\rt]
(\frac{\tau}{2},2\tau)},\\
\theta
\lt[\begin{array}{c}\frac{1}{2}\\[2pt]\frac{1}{2}\end{array}\rt]
(2u,2\tau)&=&\theta
\lt[\begin{array}{c}\frac{1}{2}\\[2pt]\frac{1}{2}\end{array}\rt]
(\tau,2\tau)\,\times\,\frac{\s(u)\s(u+\frac{1}{2})}
{\s(\frac{\tau}{2})\s(\frac{1}{2}+\frac{\tau}{2})},\\
\theta \lt[\begin{array}{c}0\\\frac{1}{2}\end{array}\rt]
(2u,2\tau)&=&\theta
\lt[\begin{array}{c}0\\\frac{1}{2}\end{array}\rt]
(0,2\tau)\,\times\,\frac{\s(u-\frac{\tau}{2})\s(u+\frac{1}{2}+\frac{\tau}{2})}
{\s(-\frac{\tau}{2})\s(\frac{1}{2}+\frac{\tau}{2})}.
\eea


\section*{Appendix B: The proof of (\ref{Id-function-1})-(\ref{Id-function-2})}
\setcounter{equation}{0}
\renewcommand{\theequation}{B.\arabic{equation}}
In this appendix, we give the proof of
(\ref{Id-function-1})-(\ref{Id-function-2}) for the eight-vertex
model. For the case of the six-vertex model, the equations  can be
proven in a similar way (or by taking the corresponding limits of
the eight-vertex case).

The $Z_2$-symmetry (\ref{Z2-sym}) of the R-matrix  implies that
\bea
 &&tr_0\lt\{(V_0K^-_0(u-\eta)V_0)^{t_0}P_{0j}R_{j0}(-2u)\rt\}\no\\
 &&\quad =tr_0\lt\{V_0\,(K^-_0(u-\eta))^{t_0}\,V_0P_{0j}V_0V_jR_{j0}(-2u)V_0V_j\rt\}\no\\
 &&\quad=-V_jtr_0\lt\{V_0\,(K^-_0(u-\eta))^{t_0}\,P_{0j}R_{j0}(-2u)V_0\rt\}V_j\no\\
 &&\quad=V_jtr_0\lt\{P_{0j}R_{j0}(-2(u-\eta)-2\eta)(K^-_0(u-\eta))^{t_0}\rt\}V_j.
\eea
The following useful relations of the K-matrices given by (\ref{K-matrix1}) and (\ref{K-matrix2})
were proven in \cite{Yan08-e} (for details we refer the reader to see (4.11), (4.12) and (A.1) in
the Ref.\cite{Yan08-e})
\bea
tr_0\lt\{(V_0K^-_0(u-\eta)V_0)^{t_0}P_{0j}R_{j0}(-2u)\rt\}&=&\frac{\s(2u-2\eta)}{\s(\eta)} K_j^-(-u),\\
tr_0\lt\{P_{0j}R_{j0}(2u)K^+_0(u)^{t_0}\rt\}&=&-\frac{\s(2u+2\eta)}{\s(\eta)} V_jK_j^+(-u-\eta)V_j,\\
K_j^-(u)\,K^-_j(-u)&=&\prod_{l=1}^3\frac{\s(\a^{(-)}_l+u)\s(\a^{(-)}_l-u)}
{\s(\a^{(-)}_l)\s(\a^{(-)}_l)}\times {\rm id}_j.
\eea The above relations give rise to
\bea
&&K_j^-(u) tr_0\lt\{(V_0K^-_0(u-\eta)V_0)^{t_0}P_{0j}R_{j0}(-2u)\rt\}=\frac{\s(2u-2\eta)}{\s(\eta)}K_j^-(u)\,K^-_j(-u)\no\\
&&\quad\quad=\frac{\s(2u-2\eta)}{\s(\eta)}\prod_{l=1}^3\frac{\s(\a^{(-)}_l+u)\s(\a^{(-)}_l-u)}
{\s(\a^{(-)}_l)\s(\a^{(-)}_l)}\times {\rm id}_j\no\\
&&\quad\quad={\rm Det}_q(K^-(u))\times {\rm id}_j,
\eea and
\bea
&&tr_0\lt\{K^+_0(u)P_{0j}R_{j0}(2u)\rt\}\,\lt\{V_jK^+_j(u-\eta)V_j\rt\}^{t_j}\no\\
&&\quad\quad= -\frac{\s(2u+2\eta)}{\s(\eta)}\prod_{l=1}^3\frac{\s(\a^{(+)}_l+u)\s(\a^{(+)}_l-u)}
{\s(\a^{(+)}_l)\s(\a^{(+)}_l)}\times {\rm id}_j\no\\
&&\quad\quad={\rm Det}_q\lt\{K^+(u)\rt\}\times {\rm id}_j. \eea
This completes the proof of  (\ref{Id-function-1})-(\ref{Id-function-2}).


\end{document}